\documentclass[sigchi]{acmart}

\AtBeginDocument{%
  \providecommand\BibTeX{{%
    \normalfont B\kern-0.5em{\scshape i\kern-0.25em b}\kern-0.8em\TeX}}}

\begin{document}

\title{Securing Manufacturing Using Blockchain}

\author{Z. Jadidi, A. Dorri,  R. Jurdak, and C. Fidge}
\affiliation{%
  \institution{Queensland University of Technology (QUT), Cyber Security Cooperative Research Centre}
}

\begin{abstract}
 Due to the rise of Industrial Control Systems (ICSs) cyber-attacks in the recent decade, various security frameworks have been designed for anomaly detection. While advanced ICS attacks use sequential phases to launch their final attacks, existing anomaly detection methods can only monitor a single source of data. Therefore, analysis of multiple security data can provide comprehensive and system-wide anomaly detection in industrial networks. In this paper, we propose an anomaly detection framework for ICSs that  consists of two stages: i) blockchain-based log management where the logs of ICS devices are collected in a secure and distributed manner,  and ii) multi-source anomaly detection where the blockchain logs are analysed using multi-source deep learning which in turn provides a system wide anomaly detection method.   
 We validated our framework using two ICS datasets: a factory automation dataset and a Secure Water Treatment (SWAT) dataset. These datasets contain physical and network level normal and abnormal traffic. The performance of our new framework is compared with single-source machine learning methods. The precision of our framework is 95\% which is comparable with single-source anomaly detectors.
 
\end{abstract}


\keywords{Industrial control systems, blockchain, anomaly detection, log management, deep learning, sequence classification}

\maketitle

\section{Introduction}\label{sec:intro}
  




With the advances in Internet and emergence of new technologies such as Internet of Things (IoT), traditional industrial control systems (ICS) are shifting from isolated sites to interconnected networks. The high degree of connectivity increases the security risks in ICS as malicious nodes may access the devices and cause malfunctions. The existing  ICSs are a collection of interconnected industrial legacy systems (operational technology (OT) networks) integrated with information technology (IT) networks \cite{ref1}. ICS networks contain  heterogeneous interconnected components such as: remote terminal units (RTU), programmable logic controllers (PLC), historian servers, and human-machine interfaces (HMI) \cite{ref7}. Studies show that the number of documented attacks targeting these ICS infrastructures by bypassing IT security has increased dramatically in recent years \cite{ref35, ref36}. This has raised the concerns for anomaly detection in ICS networks.  \par
One of the fundamental methods proposed in the literature to detect anomaly is to analyse the logs generated by ICS devices,  that potentially reflects their behaviour  and thus the actions happened in the network, using machine learning algorithms. Different supervised and unsupervised machine learning (ML) methods have been employed to detect anomalous behaviour in logs generated by ICS networks, hosts or sensors \cite{ref10, ref32, ref33, ref34}. These methods have improved the accuracy of anomaly detection in a single type of ICS logs. However, they only monitor a specific component of a network and they cannot scale to large networks with multiple sources of log generation. \par
A ML algorithm uses the logs as input to decide on anomalies, thus it is highly critical to ensure the security and integrity of the log files and to prevent logs from being deleted  accidentally or deliberately. The integrity of the logs prevents malicious nodes from changing the content of the log once it is generated by the devices. The malicious nodes may attempt to remove their trace after conducting an attack by removing the logs produced by the device. The logs are normally stored in the device itself or may be sent to a centralised cloud server. Even if the logs are sent to the cloud, the attacker can remove the logs by compromising the cloud server. In some cases, the logs might be removed accidentally by the employees of a company which in turn removes the historical behaviour of the devices that complicates the attack detection process. \par 
Different types of logs are required in ICS networks which include: a) Host  and security logs: these include logs from operating systems, firmware, and security tools, b) Network logs: these include logs that reflect communications in a network, and  c) Industrial logs: these include the logs generated by  sensors and process control equipment.  
Conventional logging systems rely on the logs produced by either a single device or devices in the same site. However, a manufacturer might have multiple sites in  geographically distributed locations which all might be the target of the same malicious behaviour. Sharing log information of all the devices in distributed locations facilitates detection of anomalous behaviours. Thus, logging in manufacturing demands a distributed, secure, auditable and immutable solution.\par 

Blockchain can potentially  address the outlined challenges due to its salient features which includes decentralisation, security, auditability and immutability \cite{ref40}. In blockchain all transactions, i.e., communications between participants, are broadcast and verified  by all participants. Particular nodes, known as validators, collect transactions and store them in blockchain in the form of blocks by following a consensus algorithm. The latter ensures blockchain security and establishes trust in the state of the ledger among untrusted participants. The transactions and blocks are publicly available to the participants which introduces high auditability. Each block is chained to the previous block in the blockchain by storing the hash of the previous block which in turn makes it impossible to modify the content of the previously stored information in the blockchain and thus introduces immutability.  \par 

Ahmed et al. \cite{ref41} proposed a blockchain-based solution for log management. The authors employed hyperledger as the underlying blockchain technology and demonstrated the blockchain applicability and feasibility of storing log information in blockchain. Similarly, Schorradt et al. \cite{ref42} studied the feasibility of using blockchain to store logging information for ICSs. The experiment shows that blockchain can potentially serve as a logging infrastructure for ICSs. The existing solutions for log management using blockchain only consider sharing data neglecting how to detect anomalous behaviour based on the collected data. \par 

Motivated by the above discussions, in this paper we develop an anomaly detection framework for ICSs. Our new framework comprises two main phases which are: 

\begin{itemize}
    \item logging that stores the log information of the devices in blockchain. The devices may be in different geographically dispersed locations within the same organisation. If any third party is involved in the process of a manufacturer, their corresponding logs is stored in the blockchain, 
    \item anomaly detection where a particular node in the network runs a multi-source deep learning algorithm to detect anomalies in logs received from multiple sources \cite{ref37}. The multi-source anomaly detector transforms the heterogeneous inputs into the same format. The output of the deep learning identifies the input source generating the anomalous behaviour. The multi-source deep learning uses Long Short Term Memory (LSTM) networks to analyse input sequences from ICS sources \cite{ref38, ref39}. To reduce the delay in detecting the anomalies, the anomaly detection is run at two levels. The first level is in each site where the site manager runs the deep learning algorithm on the logs collected from the local devices to find anomalies. The site managers in the same organisation may share their knowledge or information about any malicious activity which in turn facilitates anomaly detection in all sites. In the second level called an organisation level, a computationally capable device runs deep learning on all logs in the blockchain to detect the anomalies which in turn facilitates detecting more complicated malicious behaviours. Once a device generates a log, it immediately sends the log in the form of a transaction to its neighbouring nodes to be stored in the blockchain which prevents malicious nodes from modifying the logs. Blockchain is stored by multiple nodes which  protects against deliberately or accidentally removing the logs. 
\end{itemize}

The rest of this paper is organised as follows. Section 2 explains background and related works. Section 3 discusses the threat model used in our study. Section 4 describes the anomaly detection framework proposed in this paper. Evaluation datasets are explained in Section 5. Section 6 discusses the experimental results of the presented framework. Section 7 concludes the paper.



\section{Background and related works}\label{sec:related-work}
This study presents an anomaly detection framework for the ICS architecture provided in the Purdue model  \cite{ref2, ref3, ref4}. In this section we provide a background discussion on the existing anomaly detection methods in Section \ref{sub:sec:anomaly-detection} and blockchain in log management in Section \ref{sub:sec:BCforLog}.  \par


\subsection{Anomaly Detection in ICS}\label{sub:sec:logmannagement}

Analysis of ICS logs has been investigated in the literature. Feng and Chana \cite{ref8} combined  the content of the network packets  transmitted between ICS components and their time-series data to detect anomalies. Communication between ICS devices follow regular and predictable patterns which is used to generate a base-line of normal patterns for ICS network. A Bloom filter and a stacked LSTM network-based softmax classifier were used in their work \cite{ref8}. This combinational method can predict the future behaviour of the data created from a gas pipeline SCADA system and detect anomalies with high performance. \par
Unsupervised machine learning has been also applied to anomaly detection in ICS networks \cite{ref9}. The detection rates of Deep Neural Networks (DNN) and one-class Support Vector Machines (SVM) were compared using Secure Water Treatment (SWaT) dataset captured from a simulated fully operational raw water purification plant. These two methods were initially trained using logs of normal conditions in SWaT dataset, and then, they were evaluated by 36 attack scenarios in this dataset. DNN generates fewer false positives and better F1 measure than SVM. Convolutional Neural Networks (CNN) is another deep learning method applied in anomaly detection in ICSs \cite{ref10}. This anomaly detector has also been evaluated using the SWaT dataset. The statistical deviation of the predicted value from the observed value were measured to detect anomalies. The proposed method employed a variety of deep neural network methods including different kinds of convolutional and recurrent networks. The results show the efficiency of 1D convolutional networks in predicting time-series prediction tasks and in detecting anomalies in ICS networks. \par
Stateful anomaly detection based on the cumulative sum (CUSUM) of residuals was proposed by Ghaeini et al. \cite{ref11} to analyse physical process logs. The proposed method uses state dependent detection thresholds to control an attacker trying to manipulate ICS data process. This state-aware system has three characteristics. It can predict system state from the historical data of the network and physical process. It provides cumulative sum of residuals for monitoring ICS. Thirdly, it provides state-aware anomaly detection. The proposed method is evaluated using about 120 GB of historical data from the ICS, and it provides less time-to-detect of attacks and fewer false alarms compared to other methods. As the current system state is used by Ghaeini et al.  \cite{ref11} to compute CUSUM residuals, this method could prevent a stealthy attacker. However, the proposed method in their paper adds computation overhead as it perform pre-computation of the physical process state information.\par

While these ML-based anomaly detection methods focus on a single type of input data, our multi-source deep learning can detect anomalies in data received from multiple heterogeneous sources. The existing works in anomaly detection assume that the log data is transmitted securely and reliably without any modification on the log content, while in real-world scenarios, the attacker may attempt to remove the trace or alter the  logs to cover its track. Most of the existing works assume the devices that generate logs are in the same site while in most manufacturers the devices might be in geographically dispersed locations \cite{ref47}. Third parties may also be involved in manufacturing, e.g., service center that services devices. It is critical to maintain a comprehensive log. The log information contains privacy-sensitive information about the manufacturer and thus should be kept private from third-parties. Blockchain is employed to increase the security of log management which is studied next.

\subsection{Blockchain in Log Management} \label{sub:sec:BCforLog}
In this section we study the log management systems that are based on blockchain. Ahmad at el. \cite{ref41} proposed a blockchain-based logging framework. The proposed framework employs a private blockchain where only authorised nodes are allowed to participate in storing logs. The end devices send the logs to a server to be translated to blockchain transaction format (that is JSON in the paper). The transaction is then stored in the blockchain visible to all other participants. Byzantine Fault Tolerance (BFT) is employed as the underlying consensus algorithm to reduce the overheads associated with storing new blocks in the blockchain. \par 

Pourmajidi and Miranskyy \cite{ref43} proposed Logchain, a blockchain-assisted log storage. Logchain consists of two tiers which are: i) Circulated chain: this tier comprises of a circulated blockchain that has a genesis block, i.e., the first block in the ledger, and termination block that is the last block in the ledger and is linked back to the genesis block forming a circulated chain. The number of blocks in circulated chain must be identified initially by the system managers, and ii) super chain: this tier is the main blockchain that contains the hash of the circulated blockchain which  increases the throughput of the blockchain and reduces the associated packet and processing overheads. In a similar attempt, Tomescu and Devadas \cite{ref44} proposed Catena, where a log of a number of events is stored in blockchain to improve throughput. A logging server receives the events and generates a log once the number of events reaches a pre-defined value.\par

Castaldo and Cinque \cite{ref45} proposed a blockchain-based logging  system to store the logs related to health data exchange between countries. The framework is integrated with the existing health systems. The information from the health system is sent to a server that exchanges the data with the requestee while storing the corresponding log in the blockchain. A private blockchain is used where all participants are authorised by using CAs in their corresponding region. \par 

Rane and Dixit \cite{ref46} proposed a framework to store log data of cloud sources in the blockchain. A node controller collects log information from different sources and encrypts the log information with the public key associated with the relevant party who needs to access the data. Next, the controller sends the data to the blockchain. The intended receiver of the data collects data from the blockchain and decrypts with the  private key associated with its public key.\par 
Schorradt et al. \cite{ref42} proposed a prototype implementation of blockchain-based logging system for ICSs. The authors argue that blockchain immutability and auditability makes it attractive to address security challenges in ICS logging systems.\par

While the existing works in blockchain-based logging systems focus only on storing information in the blockchain, our solution provides a comprehensive framework to store and process log information to detect anomalous behaviour in ICS. Most of the existing solutions rely on conventional blockchains, while such blockchain management involves packet and computational overheads which are far beyond the capabilities of the devices in manufacturing industry. 

\section{Threat Model} 
In this section we discuss the threat model. The devices accept communications from the Internet as the admins may remotely control, monitor, and update the devices. Only devices authorised by ICS admins can participate in the blockchain. Logging information stored in the blockchain is not visible to all participants and just authorised nodes have read/write permissions. Secure asymmetric  encryption algorithms employed which cannot be compromised by the malicious nodes. It is assumed that the devices generate a log corresponding to an event immediately after conducting the event. The log is broadcast to the neighbor devices in the same site. \par
The device logs in each site are forwarded to a multi-source deep learning detector to monitor all devices and detect anomalies. The output of the anomaly detector will be securely shared with other sites using blockchain.  

\section{Anomaly detection in ICS}\label{sec:proposedmethod}
\begin{figure}
    \centering
    \includegraphics[scale=0.5]{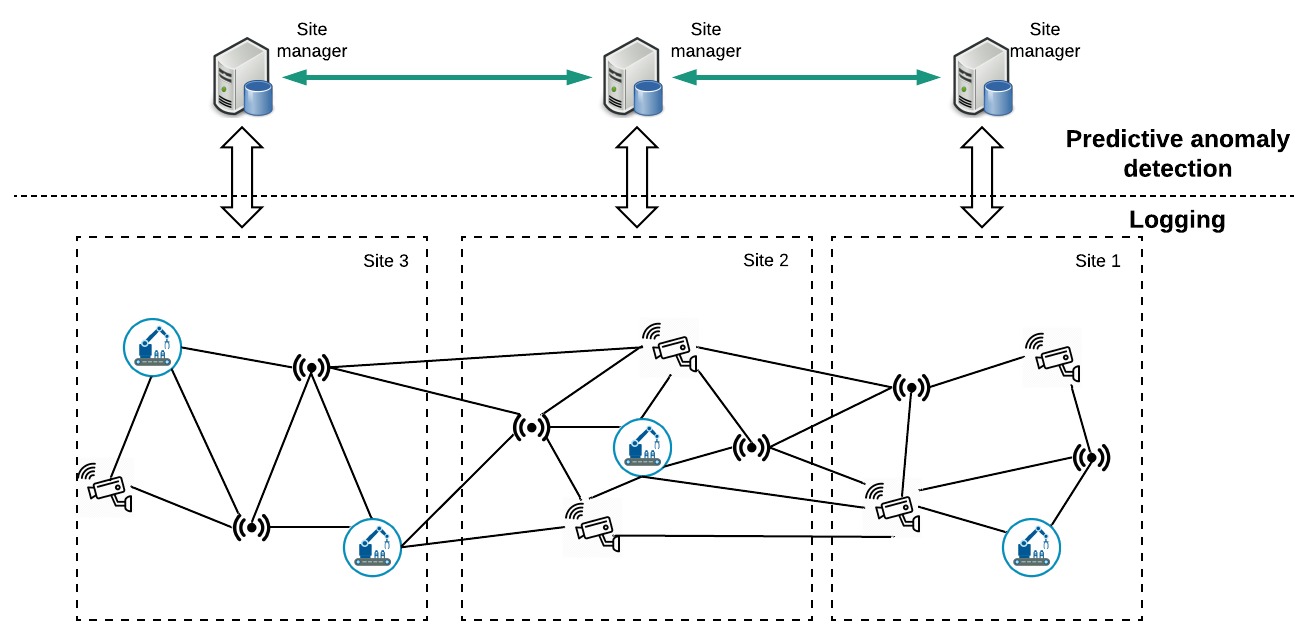}
    \caption{An overview of the proposed framework. }
    \label{fig:overview}
\end{figure}

In this section we outline the details of the proposed anomaly detection framework. The manufacturer establishes a private blockchain and authorises the devices to join the blockchain by generating a genesis transaction, i.e., the first transaction in a ledger, for each device that can be used by the device to chain its following transactions. Figure \ref{fig:overview} represents a high level view of the proposed framework. The participants in the private blockchain have different read/write permisssions in a way that only validators, i.e., the group of nodes that store the blocks in the blockchain, have write permission. The validators are authorised by the manufacturer and can be the site managers. It is possible that multiple third parties are involved in the private chain of a manufacturer to store their logging information. As the log information is privacy-sensitive to the manufacturer, it is critical to ensure that only authorised nodes can read and access such data.  The manufacturer announces a PK, known as \textit{PK\textsubscript{data}} that shall be used by the devices to encrypt log data. \textit{PK\textsubscript{data}} and the associated private key is known to the entities that require to read log data as further discussed in Section \ref{sub:sec:anomaly-detection}. \par

Blockchain is only employed as a  shared trusted database to store the logging information of the devices while the rest of the communications of the devices is happening in an off-chain channel, e.g., through the Internet. The proposed framework consists of two phases which are logging, where devices store logs in the blockchain, and predictive anomaly detection, where logs are analysed to detect malicious activities which are discussed in greater details below.

\subsection{Logging}\label{sub:sec:logging}
In this phase, the devices store their logs in the private chain. Only authorised devices that are installed by the manufacturer can participate and share information with other parties. Thus, during the bootstrapping the manufacturer must generate a genesis transaction for each  device that is structured as follows:\par 

\textit{T\_ID || PK || PK\textsubscript{data}}\par 

where \textit{T\_ID} represents the identity of the transaction which is essentially the hash of the transaction content, \textit{PK} is the public key employed by the device to generate the next transaction, and \textit{PK\textsubscript{data}} is the public key that must be used by the device to encrypt data. A manufacturer may need change \textit{PK\textsubscript{data}} later to enhance security of the data. To do so, the manufacturer generates a \textit{key update} transaction and includes the new \textit{PK\textsubscript{data}} and is signed with the private key associated with the previous \textit{PK\textsubscript{data}}. \par 

When an event occurrs, the device generates  a \textit{Log Store (LS)} transaction that is structured as follows:\par 

\textit{T\_ID || P\_T\_ID || Log || PK || Signature}\par 

where \textit{T\_ID} is as in the genesis transaction, \textit{P\_T\_ID} is the hash of the previous transaction generated by the device that creates a ledger between the transactions generated by a device. This also ensures that only the authorised nodes can participate in the blockchain. \textit{Log} is the   log of the device encrypted with \textit{PK\textsubscript{data}}. The last two fields are the \textit{PK} of the device and the corresponding signature which ensures that the transaction generator knows the private key corresponding to the \textit{PK} and thus protects against malicious nodes that may pretend to be an authorised device. The device broadcasts \textit{LS} transaction to its neighbor nodes to be stored in the blockchain.  \par

The immutability offered by  blockchain makes it impossible for the malicious entities to modify previously stored  logs which in turn increases the security of the framework. Having discussed the logging process, we next discuss predictive anomaly detection.

\subsection{Predictive anomaly detection} \label{sub:sec:anomaly-detection}
The main objective of this phase is for the site or/and manufacturer manager to detect anomalies in the network. Our framework introduces two levels of anomaly detection which are site and organisation level. In  each site a device is dedicated to run the deep learning algorithm, referred to as site manager in the rest of this paper,  to detect anomalies.  Details of the deep learning algorithm are introduced later in this section. If the site manager detects any anomaly or suspicious behaviour from a particular user, it informs other site managers in the private chain by generating an Anomaly Alert (\textit{t\textsuperscript{aa}}) that is structured as follow:

\textit{T\_ID ||P\_T\_ID  || M\_PK  ||   Log || PK || Signature  }\par 
where  \textit{M\_PK} is the \textit{PK} of the suspicious node that potentially helps other site managers to identify  suspicious node and monitor its transactions in their site. The malicious node may change their \textit{PK} in different sites  which makes it challenging to detect anomalies  only based on  \textit{PK} of the malicious node. Thus,  the site manager stores the  pattern of an anomaly in \textit{Log}  field which assists  the site managers to detect the same  pattern of actions. \par 

An anomaly might involve devices in different sites that makes it challenging for  the site managers to detect the attack as they analyse the log data of their corresponding devices. At the organisation level, the manufacturer can dedicate a device to run a deep learning algorithm on top of the blockchain data that is the data of all the devices in all sites. This  increases the chance of detecting attacks  that involve multiple devices in different sites as the volume of input data increases. \par  

While numerous supervised and unsupervised machine learning methods have been proposed for ICS security \cite{ref10, ref32, ref33, ref34}, they only focus on detection of local anomalies and do not provide system wide anomaly detection. Multi-source deep learning has attracted researchers in different data management areas \cite{ref28, ref29}. In this method, a deep learning method is trained with the time sequence data transferred from multiple sources.\par
\begin{figure}
\centerline{\includegraphics[width=18.0pc]{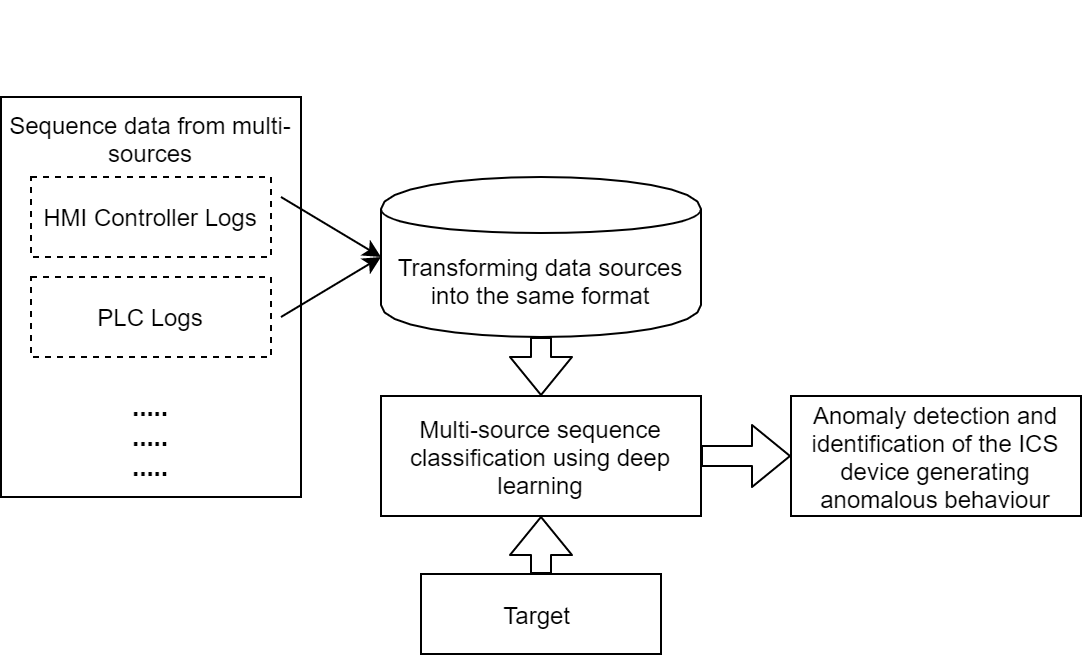}}
\caption{The architecture of anomaly detection in ICS using multi-source deep learning}
\label{fig2}
\end{figure}

 
In our framework, a multi-source deep neural network (MS-DNN) is deployed to solve the problem of anomaly detection in data received from multiple ICS devices over different Purdue layers \cite{ref24}. 
Our MS-DNN solution uses sequence classification to detect anomalies. Figure ~\ref{fig2} shows the architecture of ICS anomaly detection using MS-DNN. MS-DNN is a supervised classifier which needs a labeled dataset. These labels shown as Target in Figure ~\ref{fig2}.
The presented MS-DNN solution has multiple inputs which are transformed into the same format. The combined inputs will be a cell array maintaining the historical data of all inputs. The output of MS-DNN is the number of device logs being monitored by the anomaly detector. MS-DNN is a classifier that detect anomalies in a site level and identifies the ICS device generating the anomalies \cite{ref21, ref22}. 

\begin{figure}
\centerline{\includegraphics[width=15.0pc]{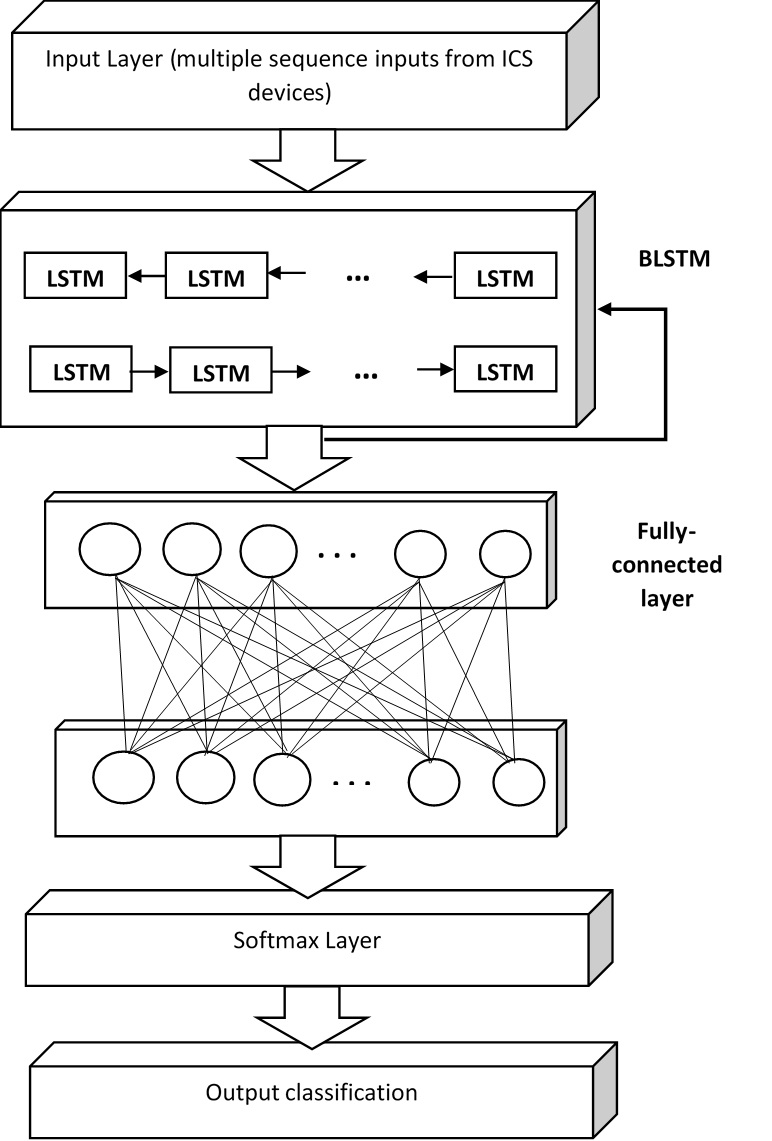}}
\caption{Design of MS-DNN layers}
\label{fig5}
\end{figure}

Deep learning neural networks are an end-to-end solution for network traffic analysis. They can extract features from raw data and use the extracted features for classification tasks. In our MS-DNN, an LSTM network is used to train a deep neural network to:
\begin{itemize}
    \item make prediction based on sequence data received from ICS devices, and 
    \item classify the sequences into devices, and identifies whether the device behaves normally. The output of the deep learning corresponds to the number of end devices. 
\end{itemize}
The design of the multi-source deep learning is shown in Figure ~\ref{fig5}. The LSTM in this deep learning has inputs from different sources. A bidirectional (BLSTM) layer with 100 hidden units is used. The LSTM layer is connected to a fully connected layer followed by a softmax layer and a classification layer  \cite{ref26, ref27}.\par
The performance of MS-DNN was evaluated based on two publicly available ICS datatsets, and its accuracy and precision were compared with other classifiers \cite{ref25}. Two MS-DNNs were separately used for anomaly detection in site manager levels. However, anomaly detection at an organisation level is not evaluated in this paper. \par

Once the manufacturer or the site manager detects an anomaly, they shall inform "incident response team" (IRT) to take further actions and secure the network. To this aim, they generate a \textit{t\textsuperscript{aa}} as discussed earlier in this section and include the relavant information about the attack. On receipt of \textit{t\textsuperscript{aa}}  IRT first verifies the signature of the transaction generator to ensure an authorised device has generated the same. Then, the data about the anomaly is analysed to decide on proper actions. The decision processing in this step is beyond the scope of this paper. IRT uses the blockchain information to verify the log files and find the liable party. Benefiting from the blockchain immutability, IRT and other authorities can trust the log information stored in blockchain.

\section{Datasets} \label{sec:implementation}

Two datasets were used to evaluate our MS-DNN solution. These datasets provide multiple ICS data sources required for training of MS-DNN.

\begin{itemize}
	\item The first dataset (SWAT dataset) is of a small scale industrial water treatment process, which is a six stage system. The dataset was generated by the iTrust cyber security research center \cite{ref30}. The communication protocol used for the automation was Modbus. The attacks on the system were spoofing and man-in-the-middle. SWAT dataset includes physical data collected from sensors and actuators and network data.
	
	\item The second dataset (factory automation dataset) was generated from a three-stage, laboratory-based factory automation process consisting of a conveyor belt sorting system, a water tank system, and a pressure vessel system. The process is automated using SIEMENS PLCs. Thus, the communication protocol is S7comm \cite{ref31}. All the attacks on the system were flooding attacks from an attacker machine on the same network. 
	
\end{itemize}

These datasets provide network traffic captured in two locations: 1) Pcap
logs captured from ICS devices like PLCs, and 2) Pcap files captured from a switch
connected to the attacker's network. Two datasets in this paper demonstrate two ICS sites, A and B, and they are used to evaluate the performance of MS-DNN based anomaly detection in site management levels. 

\section{Evaluations and Results}\label{sec:evaluation}

The MS-DNN uses LSTM networks to classify sequence data received from ICS devices. An LSTM uses sequence data as input and makes prediction based on the time steps of the sequence data. The first training dataset in this paper (SWAT dataset) contains time series for two data sources from physical devices and network traffic. Each sequence has a number of features, separately extracted from each source, and it varies in length. The maximum number of features is defined as the required number of features in all input sequences in deep learning. Twelve TCP/IP header features were manually extracted from network Pcap files. Therefore, input sequences were defined with twelve features which shows the maximum number of features in our datasets. \par
\begin{figure}
\centerline{\includegraphics[width=19.0pc]{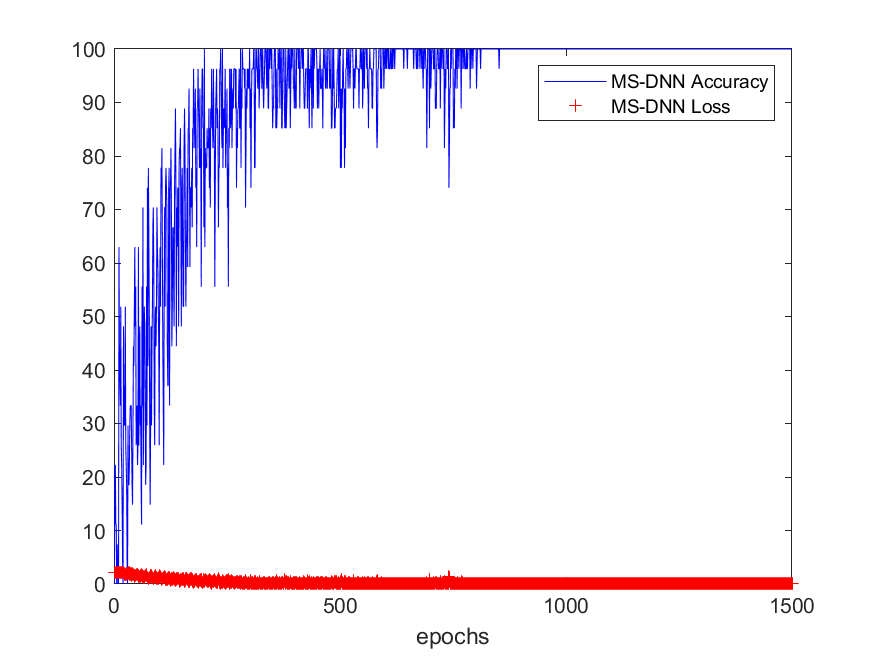}}
\caption{MS-DNN Training accuracy and errors in SWAT dataset}
\label{fig3}
\end{figure}
\begin{figure}
\centerline{\includegraphics[width=19.0pc]{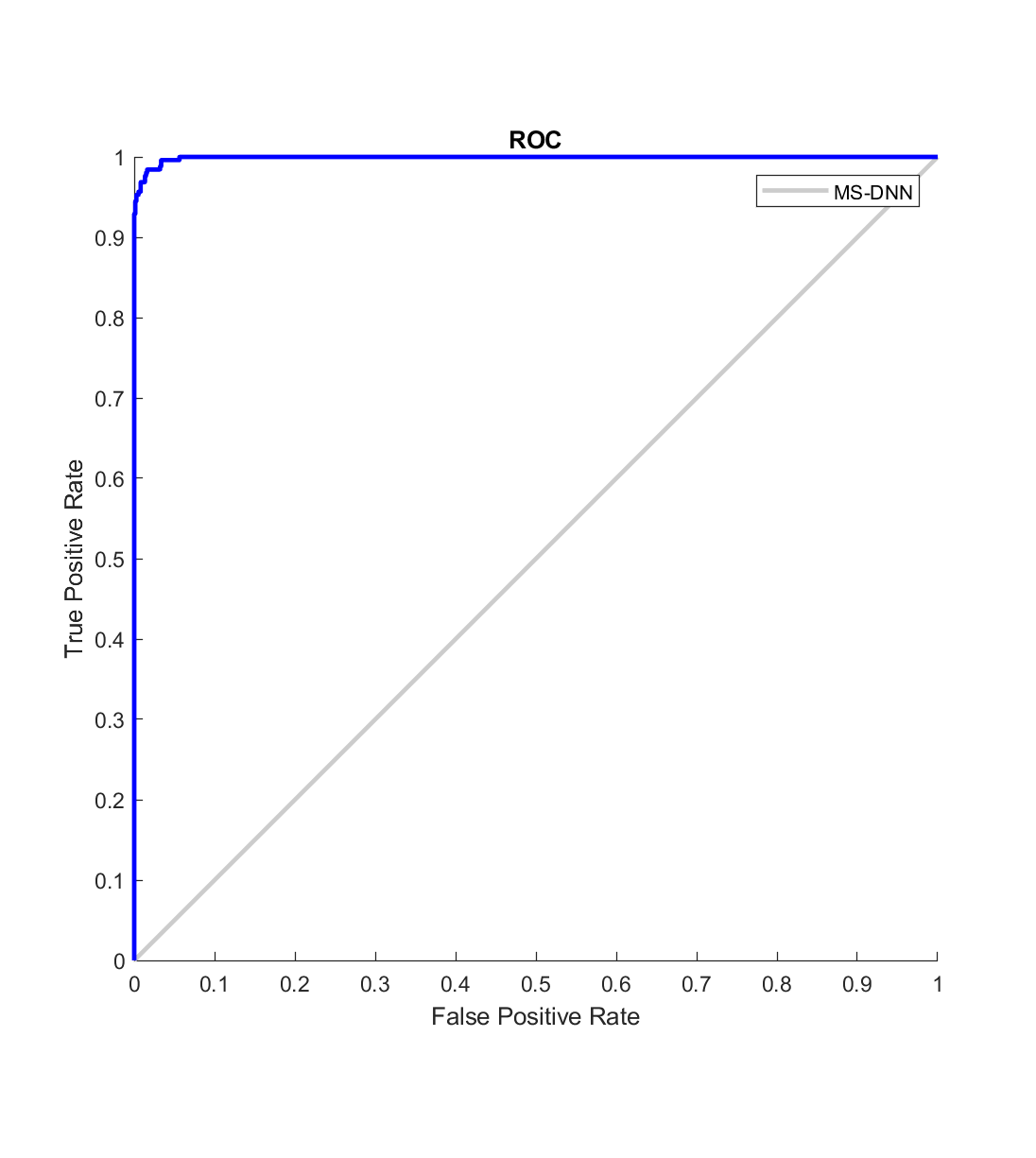}}
\caption{ROC curve for MS-DNN in SWAT dataset}
\label{fig6}
\end{figure}

The training dataset contains 2000 samples and testing dataset has 1000 samples. The output detects anomalies in input sequences and identifies the device generating the anomalous behaviour \cite{ref23}. Figure ~\ref{fig3} shows the accuracy of MS-DNN with different iterations in the training phase in SWAT dataset. As this figure illustrates, the accuracy is 100\% after 1000 iteration. ROC curve of the MS-DNN,  Figure ~\ref{fig6}, also shows the high True Positive rate of this anomaly detector. \par

The F1 score is the performance metric used in this paper, as shown in Eq. \ref{eq:1} \cite{ref10}. The MS-DNN was separately evaluated using SWAT dataset and Factory automation dataset, and its results are shown in Table~\ref{tab:table1}.

\begin{equation} \label{eq:1}
F_{1}=2.\frac{precision\cdot recall}{precision+recall}
\end{equation}

Table~\ref{tab:table1} shows the experimental results of the implemented MS-DNN in comparison with other studies \cite{ref10}. As it is shown, while MS-DNN analyses multiple inputs, the precision and F1 metrics are still similar to other machine learning based anomaly detectors with a single-source input. According to the Purdue model, ICS networks have a five-layer structure with different types of devices. For anomaly detection using single source analysis, a machine learning method is required for each type of data. However, the MS-DNN is trained by the historical behaviour of multiple ICS devices. Sequential dependency between input data points are used for sequential classification tasks. LSTM in the MS-DNN can learn long term dependencies in sequences and this helps to detect the anomalous device in input sources. 
Therefore, MS-DNN receives sequence data from input sources and classify the data into normal / abnormal classes. It also identifies the source generating anomalies.  

\begin{table}[h!]
\caption{Validation results for MS-DNN}
\label{tab:table1}
\resizebox{\columnwidth}{!}{%
\begin{tabular}{ c c c c c c }
\hline
\tiny Dataset & \tiny Method & \tiny Precision & \tiny Recall & \tiny F1  & \tiny Accuracy \\
\hline
\tiny Factory automation dataset & \tiny MS-DNN & \tiny 0.96 &  \tiny 0.80 &	\tiny 0.87  & \tiny 0.97 \\
\hline
\tiny SWAT & \tiny MS-DNN & \tiny 0.95 & \tiny 0.82 & \tiny 0.88  & \tiny 0.95 \\
\hline
\tiny SWAT & \tiny SVM \cite{ref10} & \tiny 0.93 & \tiny 0.70 &	\tiny 0.80  & \tiny ... \\
\hline
\tiny SWAT & \tiny 1D CNN combined records \cite{ref10} & \tiny 0.97 &  \tiny 0.79 &	\tiny 0.87  & \tiny ... \\
\hline
\tiny SWAT & \tiny 1D CNN ensembled records \cite{ref10} & \tiny 0.87 &  \tiny 0.85 &	\tiny 0.86  & \tiny ... \\
\hline
\end{tabular}
}
\end{table}

\section{Conclusion and Future works} \label{sec:conclusion}
In the present work, we provided an anomaly detection framework for ICS networks. This framework used blockchain to securely store device logs. In addition, multi-source deep learning was used to detect anomalies at two levels, the site and organisation levels. Site-level anomaly detection was implemented to analyse device logs in each site. Then, organisation level shared the knowledge of malicious activity in each site with other sites. \par
Once a manufacturer detects anomalous behaviour it can inform other manufacturers of the detected attack through the threat intelligence layer. In this layer, all manufacturers and other involved parties are communicating to share anomaly signs. A public blockchain is employed in this layer that increases  trust among the participants. This layer also enables the participants to generate attack signatures and share which in turn enhances the security of the framework and reduces the delay involved in generating signatures. 
This method can also be used to verify and authenticate communications as only registered devices here can participate in the main chain. Threat intelligence will be investigated in our future work. \par

Threat intelligence provides technical indicators of compromises (IoCs) and actionable advice about emerging threats. A real-time threat intelligence sharing can help to distribute knowledge about threat agents and targeted assets across individual defenders \cite{ref16, ref18}. The future aim of this research is using blockchain for secure threat intelligence. \par
While threat intelligence is a helpful technique to improve the security of a society, participants are often reluctant to share their threat information. Threat intelligence sharing using blockchain can help all participants to share their relevant information dynamically and overcome the lack of trust in threat information sharing. 
In addition, a sharing strategy based on trust and anonymity is required to help participants to share their IoCs and information about existing threats without any concern about the risk of business leak \cite{ref16, ref17, ref18}. These are the challenges which will be investigated in future work.

\subsection{Acknowledgments}
The authors acknowledge the support of the Commonwealth of Australia and Cybersecurity Research Centre Limited.

\bibliographystyle{ieeetr}
\bibliography{sample-base,mend_references}

\begin{thebibliography}{1}
\bibitem{ref1}
https://cdn.powermag.com/wp-content/uploads/2018/03/dragos\_2017-hunting-and-responding-to-industrial-intrusions.pdf
\bibitem{ref7}
Tuptuk, N., \& Hailes, S. (2018). Security of smart manufacturing systems. Journal of manufacturing systems, 47, 93-106.
\bibitem{ref35}
[2] Team, U. I. C. E. R. (2016). Recommended Practice: Improving
Industrial Control Systems Cyber security with Defense-In-Depth Strategies. Retrieved from: www.ics-cert.uscert.gov/sites/default/les/recommended practices/NCCIC ICSCERT Defense in Depth 2016 S508C.pdf
\bibitem{ref36}
[3] Hu, Y., Yang, A., Li, H., Sun, Y., \& Sun, L., A survey of intrusion detection
on industrial control systems. International Journal of Distributed Sensor
Networks. 1-14 (2018).
\bibitem{ref10}
Kravchik, M., \& Shabtai, A. (2018, January). Detecting cyber attacks in industrial control systems using convolutional neural networks. In Proceedings of the 2018 Workshop on Cyber-Physical Systems Security and PrivaCy (pp. 72-83).
\bibitem{ref32}
Justin M Beaver, Raymond C Borges-Hink, and Mark A Buckner. 2013. An evaluation
of machine learning methods to detect malicious SCADA communications.
In Machine Learning and Applications (ICMLA), 2013 12th International Conference
on, Vol. 2. IEEE, 54–59.
\bibitem{ref33}
Raymond C Borges Hink, Justin M Beaver, Mark A Buckner, Tommy Morris, Uttam Adhikari, and Shengyi Pan. 2014. Machine learning for power system disturbance and cyber-attack discrimination. In Resilient Control Systems (ISRCS)
\bibitem{ref34}
Jun Inoue, Yoriyuki Yamagata, Yuqi Chen, Christopher M Poskitt, and Jun Sun.
2017. Anomaly detection for a water treatment system using unsupervised
machine learning. arXiv preprint arXiv:1709.05342 (2017).
\bibitem{ref40}
Dorri, A., Kanhere, S. S., Jurdak, R., \& Gauravaram, P. (2019). LSB: A Lightweight Scalable Blockchain for IoT security and anonymity. Journal of Parallel and Distributed Computing, 134, 180-197.

\bibitem{ref41}
Ahmad, A., Saad, M., Bassiouni, M., \& Mohaisen, A. (2018, November). Towards blockchain-driven, secure and transparent audit logs. In Proceedings of the 15th EAI International Conference on Mobile and Ubiquitous Systems: Computing, Networking and Services (pp. 443-448).

\bibitem{ref42}
Schorradt, S., Bajramovic, E., \& Freiling, F. (2019, November). On the feasibility of secure logging for industrial control systems using blockchain. In Proceedings of the Third Central European Cybersecurity Conference (pp. 1-6).
\bibitem{ref37}
Ouyang, W., Chu, X., \& Wang, X. (2014). Multi-source deep learning for human pose estimation. In Proceedings of the IEEE Conference on Computer Vision and Pattern Recognition (pp. 2329-2336).
\bibitem{ref38}
 Jurgovsky, J., Granitzer, M., Ziegler, K., Calabretto, S., Portier, P. E., He-Guelton, L., \& Caelen, O. (2018). Sequence classification for credit-card fraud detection. Expert Systems with Applications, 100, 234-245.
\bibitem{ref39}
Kolosnjaji, B., Zarras, A., Webster, G., \& Eckert, C. (2016, December). Deep learning for classification of malware system call sequences. In Australasian Joint Conference on Artificial Intelligence (pp. 137-149). Springer, Cham.
\bibitem{ref2}
SANS Institute: Reading Room - Industrial Control Systems / SCADA, https://www.sans.org/reading-room/whitepapers/ICS/secure-architecture-industrial-control-systems-36327
\bibitem{ref3}
Automation, R. (2011). Converged Plantwide Ethernet (CPwE) Design and Implementation Guide.”.
\bibitem{ref4}
NCCIC (2016). Recommended practice: Improving industrial control systems cybersecurity with defence-in-depth strategies. US-CERT Defence In Depth, https://www.us-cert.gov/sites/default/files/recommended\_practices/NCCIC\_ICS-CERT\_Defense\_in\_Depth\_2016\_S508C.pdf
\bibitem{ref6}
https://collaborate.mitre.org/attackics/index.php/Main\_Page

\bibitem{ref8}
Feng, C., Li, T., \& Chana, D. (2017, June). Multi-level anomaly detection in industrial control systems via package signatures and LSTM networks. In 2017 47th Annual IEEE/IFIP International Conference on Dependable Systems and Networks (DSN) (pp. 261-272). IEEE.
\bibitem{ref9}
Inoue, J., Yamagata, Y., Chen, Y., Poskitt, C. M., \& Sun, J. (2017, November). Anomaly detection for a water treatment system using unsupervised machine learning. In 2017 IEEE International Conference on Data Mining Workshops (ICDMW) (pp. 1058-1065). IEEE.

\bibitem{ref11}
Ghaeini, H. R., Antonioli, D., Brasser, F., Sadeghi, A. R., \& Tippenhauer, N. O. (2018, April). State-aware anomaly detection for industrial control systems. In Proceedings of the 33rd Annual ACM Symposium on Applied Computing (pp. 1620-1628).
\bibitem{ref12}
Schorradt, S., Bajramovic, E., \& Freiling, F. (2019, November). On the feasibility of secure logging for industrial control systems using blockchain. In Proceedings of the Third Central European Cybersecurity Conference (pp. 1-6).
\bibitem{ref13}
Rafael Accorsi. 2009. Log Data as Digital Evidence: What Secure Logging Protocols Have to Offer?. In 2009 33rd Annual IEEE International Computer Software
and Applications Conference. IEEE, 398–403. https://doi.org/10.1109/COMPSAC.
2009.166
\bibitem{ref14}
Rafael Accorsi. 2010. BBox: A Distributed Secure Log Architecture. In Proceedings of the 7th European Conference on Public Key Infrastructures, Services
and Applications (EuroPKI’10). Springer-Verlag, Berlin, Heidelberg, 109–124.
http://dl.acm.org/citation.cfm?id=2035155.2035166
\bibitem{ref15}
Mihir Bellare and Bennet S. Yee. 1997. Forward Integrity For Secure Audit Logs.
Available: http://citeseerx.ist.psu.edu/viewdoc/summary?doi=10.1.1.28.7970.
\bibitem{ref16}
Tounsi, W., \& Rais, H. (2018). A survey on technical threat intelligence in the age of sophisticated cyber attacks. Computers \& security, 72, 212-233.
\bibitem{ref17}
Wagner, C., Dulaunoy, A., Wagener, G., \& Iklody, A. (2016, October). Misp: The design and implementation of a collaborative threat intelligence sharing platform. In Proceedings of the 2016 ACM on Workshop on Information Sharing and Collaborative Security (pp. 49-56).
\bibitem{ref18}
Riesco, R., Larriva-Novo, X., \& Villagra, V. A. (2019). Cybersecurity threat intelligence knowledge exchange based on blockchain. Telecommunication Systems, 1-30.
\bibitem{ref19}
Ahrend JM, Jirotka M, Jones K. On the collaborative practices of cyber threat intelligence analysts to develop and utilise tacit
Threat and Defence Knowledge, in: Cyber Situational Awareness, Data Analytics And Assessment (CyberSA), 2016 International Conference On, IEEE, pp. 1–10; 2016.
\bibitem{ref20}
OSINT, ATP 22-2.29. Open source intelligence headquarters departments of the army. Retrieved June 1, 2019 from https://fas.org/irp/doddir/army/atp2-22-9.pdf.
\bibitem{ref21}
Insider Threat Identification Using the Simultaneous Neural Learning of Multi-Source Logs.
\bibitem{ref22}
Learning from Multiple Sources
\bibitem{ref23}
Jurgovsky, J., Granitzer, M., Ziegler, K., Calabretto, S., Portier, P. E., He-Guelton, L., \& Caelen, O. (2018). Sequence classification for credit-card fraud detection. Expert Systems with Applications, 100, 234-245.
\bibitem{ref24}
Li, J., Wu, W., Xue, D., \& Gao, P. (2019). Multi-Source Deep Transfer Neural Network Algorithm. Sensors, 19(18), 3992.
\bibitem{ref25}
A Deep Learning Approach to DNA Sequence Classification
\bibitem{ref26}
A novel wavelet sequence based on deep bidirectional LSTM network model for ECG signal classification
\bibitem{ref27}
https://au.mathworks.com/help/deeplearning/ug/classify-sequence-data-using-lstm-networks.html
\bibitem{ref28}
Sun, S.; Shi, H.; Wu, Y. A survey of multi-source domain adaptation. Inf. Fusion 2015, 24, 84–92.
\bibitem{ref29}
Wu, Q.; Zhou, X.; Yan, Y.; Wu, H.; Min, H. Online transfer learning by leveraging multiple source domains. Knowl. Inf. Syst. 2017, 52, 687–707. 
\bibitem{ref30}
Goh, J., Adepu, S., Junejo, K. N., \& Mathur, A., A Dataset to Support Research in the Design of Secure Water Treatment Systems. In International Conference on Critical Information Infrastructures Security. 88-99 (2017)
\bibitem{ref31}
Myers, D., Suriadi, S., Radke, K., \& Foo, E. (2018). Anomaly detection for industrial control systems using process mining. Computers \& Security, 78, 103-125.

\bibitem{ref43}

Pourmajidi, W., \& Miranskyy, A. (2018, July). Logchain: blockchain-assisted log storage. In 2018 IEEE 11th International Conference on Cloud Computing (CLOUD) (pp. 978-982). IEEE.

\bibitem{ref44}
Tomescu, A., \& Devadas, S. (2017, May). Catena: Efficient non-equivocation via bitcoin. In 2017 IEEE Symposium on Security and Privacy (SP) (pp. 393-409). IEEE.

\bibitem{ref45}
Castaldo, L., \& Cinque, V. (2018, February). Blockchain-based logging for the cross-border exchange of ehealth data in europe. In International ISCIS Security Workshop (pp. 46-56). Springer, Cham.

\bibitem{ref46}
Rane, S., \& Dixit, A. (2019, January). BlockSLaaS: Blockchain assisted secure logging-as-a-service for cloud forensics. In International Conference on Security \& Privacy (pp. 77-88). Springer, Singapore.
\bibitem{ref47}
Drias, Z., Serhrouchni, A., \& Vogel, O. (2015, August). Analysis of cyber security for industrial control systems. In 2015 International Conference on Cyber Security of Smart Cities, Industrial Control System and Communications (SSIC) (pp. 1-8). IEEE.

\end{thebibliography}

\end{document}